\documentclass[aps,prbbib,twocolumn,epsf]{revtex4}

\newcommand{\rr}{\left({\bf r}\right)}

\usepackage{graphicx}

\begin{document}
\draft

\title{Effective medium theory of binary thermoelectrics.}
\author{Paul M. Haney$^{1}$}

\affiliation{$^{1}$Center for Nanoscale Science and Technology,
National Institute of Standards and Technology, Gaithersburg,
Maryland 20899-6202, USA }

\begin{abstract}
The transport coefficients of disordered media are analyzed using
direct numerical simulation and effective medium theory.  The
results indicate a range of materials parameters for which disorder
leads to an enhanced power factor. This increase in power factor is
generally not accompanied by an increase in the figure of merit
$ZT$, however.  It is also found that the effective electrical
conductivity and electronic contribution to the thermal conductivity
are not generally proportional to each other in the presence of
disorder.

\end{abstract}

\maketitle
\section{introduction}

There has been considerable recent interest in utilizing
nanostructure and inhomogeneity to enhance thermoelectric
performance.  Nanostructure has been predicted to reduce phonon
thermal conductivity without reducing electrical conductivity
\cite{venk}, enhance the power factor via quantum confinement
\cite{harman}, and provide strongly energy-dependent electron
scattering \cite{leonard, moyzhes, martin} - beneficial features for thermoelectrics.
One type of nanostructure consists of the inclusion of metallic
structures in a thermoelectric ({\it i.e} semiconductor) framework.
These types of nanostructured materials can exhibit two unusual
properties: unexpectedly high Seeback coefficient \cite{heremans}, and
anomalous magnetoresistance \cite{xu}.  The anomalous
magnetoresistive properties can be understood as a consequence of
disordered transport in a random resistor network model
\cite{littlewood}.

Motivated by these models of magnetoresistance, I revisit similar
models of thermoelectric properties \cite{cohen} to see if the same
mechanism can explain both phenomena.  The work presented here
extends previous studies in two ways: 1. The electrical conductivity
and Seeback coefficient of the constituent materials are varied
independently. It's found that in certain parameter regimes the
thermoelectric power factor of the composite medium is enhanced
relative to that of the constituent materials. 2. The electron and
phonon contributions to the total thermal conductivity are
calculated separately.  Although the constituent materials are
assumed to be obey the Weidemann-Franz law (W-F), the composite
medium does not. Finally, in accordance with previous studies
\cite{levy}, the figure of merit $ZT$ of the composite medium is
found to be smaller than that of the high-$ZT$ constituent material.
These conclusions are illustrated with numerical results and
analytic expressions derived within effective medium theory (EMT).

\section{Model description}
The starting point is the linear response description of transport
for the electrical current $j$ and thermal current $j_Q$:
\begin{eqnarray}
j &=& -\sigma\left({\bf r}\right) \nabla V + L^{12}\rr\nabla T ~,\nonumber \\
j_Q &=& -\left(\kappa_e\left({\bf r}\right)+\kappa_\gamma\left({\bf
r}\right)\right) \nabla T + L^{12}\left({\bf r}\right)T \nabla
V~,\label{eq:transport}
\end{eqnarray}
where $\sigma\rr$ is the local electrical conductivity,
$\kappa_e'~\left(\kappa_\gamma'\right)$ is the electron (phonon)
contribution to the total local thermal conductivity $\kappa\rr$
($\kappa\rr=\kappa_e'\rr+\kappa_\gamma'\rr$) (all thermal
conductivities evaluated for zero electric field), $V$ is the
electrostatic potential, and $T$ is the temperature.  The local
Seeback coefficient $S\rr$ is related to $\sigma\rr$ and $L^{12}\rr$ by:
$L^{12}\rr = S\rr\sigma\rr$.
I assume that $\sigma\rr$ and $\kappa_e\rr$ obey
the W-F law: $\kappa_e\rr=\sigma\rr L_0T$, where $L_0$ is the Lorenz
number. As shown in Ref. (\onlinecite{cohen}), the effective medium
electrical conductivity $\bar\sigma$, total thermal conductivity
$\bar\kappa$, and $\bar L^{12}$ satisfy:
\begin{eqnarray}
\left\langle\frac{\sigma\rr-\bar\sigma}{\sigma\rr+2\bar\sigma}\right\rangle
=0~;&&\left\langle\frac{\kappa\rr-\bar\kappa}{\kappa\rr+2\bar\kappa}\right\rangle
=0~,\label{eq:kappatotemt}
\end{eqnarray}
\begin{eqnarray}
&\bar{L}^{12}=&3\bar\sigma\bar\kappa \left\langle
\frac{L^{12}\rr}{\left(\kappa\rr+2\bar\kappa
\right)\left(\sigma\rr+2\bar\sigma\right)}\right\rangle\times
\nonumber
\\ &&\left(\left\langle\frac{\sigma\rr\bar\kappa+\bar\sigma\kappa\rr+2\bar\sigma\bar\kappa-\sigma\rr\kappa\rr}{\left(\kappa\rr+2\bar\kappa
\right)\left(\sigma\rr+2\bar\sigma\right)}\right\rangle\right)^{-1}.\label{eq:pemt}
\end{eqnarray}
The brackets indicate an average over disorder configurations.  In
addition, the electron and phonon parts of the thermal conductivity
of the effective medium, $\bar\kappa_e$ and $\bar\kappa_\gamma$,
satisfy:
\begin{eqnarray}
\left\langle\frac{\kappa_{e,\gamma}\rr-\bar\kappa_{e,\gamma}}{\kappa\rr+2\bar\kappa}\right\rangle
&=&0~,\label{eq:kappaemt}
\end{eqnarray}
Two-component materials are considered here, so that local material
parameters may assume one of two possible values.  I further suppose
that $\sigma_1 \ll \sigma_2$, and $\kappa_{\gamma1,2} \ll
\kappa_{e2}$.

The figure of merit $ZT$ can be expressed with the Onsager number $N$
(note that $N$ is constrained thermodynamically to be less than 1
\cite{davidson,mahan}).
\begin{eqnarray}
N&=& \frac{\left(\bar{L}^{12}\right)^2T}{\bar\sigma\bar\kappa_e} ~,\\
ZT &=&
\frac{N}{1-N+\left(\frac{\bar\kappa_\gamma}{\bar\kappa_e}\right)}~.
\label{eq:zt}
\end{eqnarray}

In addition to the analysis of EMT, Eq. (\ref{eq:transport}) and the
continuity equations for heat and charge ($\nabla\cdot j = 0$,
$\nabla\cdot j_Q = 0$ \cite{footnote}) are solved directly for an
ensemble of randomly disordered configurations in 3-d (correlated
disorder does not change the results appreciably).
The system is discretized into $30^3$ sites,
and the ensemble size is chosen such that the statistical error of
the effective transport parameters is converged (this typically
requires about 30 configurations).  The error bars on the plots of
numerical results indicate the statistical uncertainty (one standard
deviation).

\section{Results}
{\it Power factor} - Figs. 1(a), 1(b), and 1(c) show
$\bar\sigma,~\bar{L}^{12}$, and the power factor
$\bar{S}^2\bar{\sigma}$, respectively, as a function of
concentration of material 1 (denoted by $c$).  For this calculation
$L^{12}_1=L^{12}_2$.  At the percolation threshold ($c=2/3$),
$\bar\sigma$ shows a well-known kink \cite{kirkpatrick}, while
$\bar{L}^{12}$ is maximized. The origin of this enhancement in
$\bar{L}^{12}$ is discussed in the next section. The enhancement in
$\bar{L}^{12}$ leads to a peak in the the power factor
$\bar{S}^2\bar\sigma$ near the percolation threshold.  The figure
shows good agreement between EMT and numerical results, indicating
the EMT captures the essential physics of the power factor
enhancement. Fig. 1(d) shows the Seeback coefficient versus concentration
when the correct effective medium $L^{12}$ value is used to
calculate $S$, and when it is assumed that $L^{12}$ of the composite
medium is the same as that of the constituent materials.

\begin{figure}[h!]
\begin{center}
\vskip 0.2 cm
\includegraphics[width=3.6in]{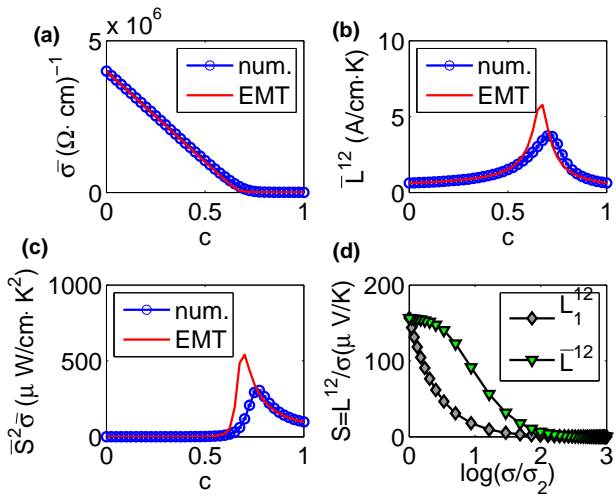}
\vskip 0.2 cm \caption{(a), (b), (c) show the effective
conductivity, $L^{12}$, and power factor as a function of disorder
concentration. (d) shows the Seeback coefficient $S=P/\sigma$ as a function
of conductivity (note the log scale of conductivity), using
$L^{12}=\bar L^{12},~L^{12}=L^{12}_1$.  Model parameters are:
$\sigma_1=4\times10^5~ {\rm \frac{1}{\Omega \cdot m}}$, $f_s=10^3$,
$\kappa_{1\gamma} = 3~{\rm\frac{ W}{m\cdot K}}$, $\kappa_{2\gamma} =
1~{\rm\frac{ W}{m\cdot K}}$,
$L^{12}_1=L^{12}_2=\sqrt{L_0}\sigma_1=62.5~{\rm \frac{A}{m\cdot
K}}$. }\label{fig:chi}
\end{center}
\end{figure}

To explain the peak in power factor, I solve Eqs.
(\ref{eq:kappatotemt}-\ref{eq:pemt}) at the percolation transition
($c=2/3$), and expand the solution in the small parameters:
\begin{eqnarray}
f_s=\frac{\sigma_1}{\sigma_2};&&f_k=\frac{\kappa_1}{\kappa_2}.
\end{eqnarray}
Keeping only leading order terms leads to:
\begin{eqnarray}
\bar\sigma&=&\sqrt{\frac{\sigma_1\sigma_2}{2}} = \sigma_1 \sqrt{\frac{1}{2 f_s}}+\ldots\label{eq:sigma}\\
\bar {L}^{12}&=&
L^{12}_1\left(\sqrt{2f_k}+\sqrt{2f_s}\right)^{-1}+\ldots\label{eq:P}
\end{eqnarray}
Eq. (\ref{eq:P}) is valid when $L^{12}_2$ is not significantly (e.g
more than 100 times) greater than $L^{12}_1$.   Eqs.
(\ref{eq:sigma}) and(\ref{eq:P}) show that both $\bar\sigma$ and
$\bar L^{12}$ diverge as $f_s^{-1/2}$ at the percolation transition.
This implies that the power factor ($\left(\bar
L^{12}\right)^2/\bar\sigma$) also diverges as $f_s^{-1/2}$ at this
point, while the Seeback coefficient ($\bar L^{12}/\bar\sigma$) remains
bounded within EMT.  The divergence of transport parameters at the
percolation transition signals a failure of effective medium theory
in the limit $f_s\rightarrow 0$.  However the numerics indicate that
EMT is sufficient for realistic parameters, and can illustrate the
key points in this work.

\begin{figure}[h!]
\begin{center}
\vskip 0.2 cm
\includegraphics[width=3.5in]{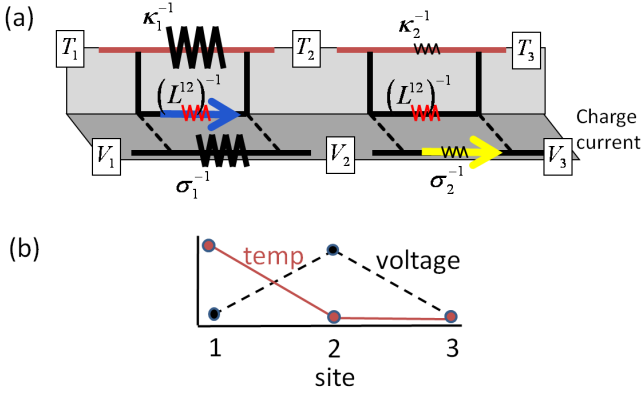}
\vskip 0.2 cm \caption{Cartoon indicating the underlying mechanism
for the enhancement of $L^{12}$ for an inhomogeneous systems
consisting of 2 resistors. (a) shows the predominant paths for charge current, (b) shows
the voltage and temperature profiles for an applied temperature
difference.
}\label{fig:resistors}
\end{center}
\end{figure}

To develop a simple picture for how inhomogeneity leads to an
enhancement in $L^{12}$, I consider Eq. (\ref{eq:transport}) for two
dissimilar materials placed in series (see Fig.
(\ref{fig:resistors})), where
$\sigma_1\ll\sigma_2,~\kappa_1\ll\kappa_2,$ and $L^{12}_1=L^{12}_2$.
$L^{12}$ of the composite system is given by $\frac{j}{\Delta
T}|_{\Delta V=0}$. The large thermal resistance of material 1
implies the temperature drop occurs mostly between sites 1 and 2.
This drives a large thermoelectric charge current between sites 1
and 2, which in turn induces a voltage at site 2.  This voltage
drives an Ohmic charge current between sites 2 and 3.  This Ohmic
current, derived from the inhomogeneous voltage, is the source of
enhanced charge current (and therefore enhanced $L^{12}$) of the
composite system. The same scenario occurs in 3-dimensional
disordered networks of materials (under the same parameter region
described above), so that when the system is maximally disordered at
the percolation transition, $L^{12}$ shows the maximum enhancement.

\begin{figure}[h!]
\begin{center}
\vskip 0.2 cm
\includegraphics[width=3.5in]{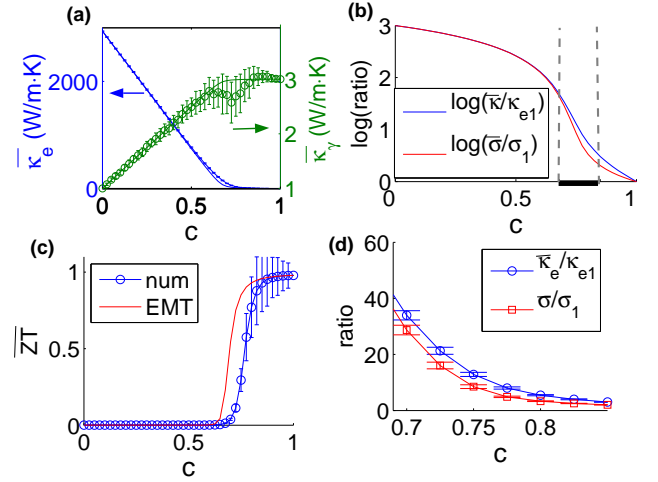}
\vskip 0.2 cm \caption{(a) shows the electron and phonon
contributions to the total thermal conductivity as a function of
concentration of material 1 (this concentration is denoted by $c$).  Solid lines are EMT result, symbols
are numerical results.  (b) shows the electrical conductivity and
electron contribution to thermal conductivity (scaled by the value
of material 1 versus concentration (note log scale).  Vertical
dashed lines indicate region that is plotted in (d). (c) shows the
figure of merit $ZT$ as a function of concentration. It generally
decreases from the high material value of $ZT=1$. (d) is a zoom-in
of (b) near percolation (on a linear scale), showing the deviation
from the W-F law of the composite medium }\label{fig:kappa}
\end{center}
\end{figure}

{\it Thermal conductivity} - Next I consider the electronic and
phonon contributions to the thermal conductivity of the disordered
material.  Let $\kappa_{\gamma 1}>\kappa_{\gamma 2}$; this describes
a system in which one material has good electronic thermoelectric
properties (i.e. a large Onsager number), but a detrimentally large
$\kappa_\gamma$, while the other material has a low Onsager number
and a small $\kappa_\gamma$.  I assume that both
$\frac{\kappa_{\gamma1}}{\kappa_{e2}}$ and
$\frac{\kappa_{\gamma2}}{\kappa_{e2}}$ are small (of the same order
as $\frac{\kappa_{e1}}{\kappa_{e2}}$).  Fig. (\ref{fig:kappa}a)
shows $\kappa_e$ and $\kappa_\gamma$ as a function of concentration.
There is good agreement between numerical results and EMT.  Fig.
(\ref{fig:kappa}b) shows that near the percolation threshold, the
electronic contribution to the thermal conductivity is not related
to the electrical conductivity via the W-F law.  At the percolation
threshold, the expressions for the total thermal conductivity
$\kappa$ and its partition into electronic and phonon parts
$\kappa_e,\kappa_\gamma$ are given as (to linear order in
$f_k,f_s$):
\begin{eqnarray}
\bar\kappa&=&\kappa_2\left[\sqrt{\frac{f_k}{2}}+\frac{f_k}{4}+\ldots\right]\\
\bar\kappa_e&=&\kappa_{2,e}\left[\sqrt{\frac{f_k}{2}}+\left(f_s-\frac{3f_k}{4}\right)+\ldots\right]\\
\bar\kappa_\gamma&=&
\kappa_{1,\gamma}+\ldots
\end{eqnarray}
The ratio $\sigma/\kappa_e$ is easily expressed in terms of
$r=f_k/f_s$:
\begin{eqnarray}
\frac{\bar\sigma}{\bar\kappa_e} = L_0 T \frac{1}{\sqrt{r}}~.
\end{eqnarray}
This indicates that in inhomogeneous materials, inferring $\kappa_e$
from $\sigma$ may not always be appropriate near the percolation
transition.  This is because the effective $\kappa_e$ is a
convolution of the intrinsic material properties and the geometry of
the system: The parallel conduction of heat current through both
electron and phonon channels (in contrast to the charge current)
results in different conducting paths for $j_Q$ and $j$, and
differences in the spatial structure of temperature and potential
fields.  As a result, the electronic part of $j_Q$ in the
inhomogeneous system is not exactly correlated with $j$.

{\it Figure of merit} - The electronic and phonon contributions to
$ZT$ can now be assembled. At percolation, the Onsager number and
ratio of electronic to phonon thermal conductivity are:
\begin{eqnarray}
N&=& N_1 \frac{1}{\sqrt{r}\left(1+r+2\sqrt{r}\right)} \\
\frac{\bar\kappa_\gamma}{\bar\kappa_e} &=&\sqrt{\frac{2}{f_k}}\left(
\frac{\kappa_{\gamma,1}}{\kappa_{e,2}} \right)
\end{eqnarray}
Plugging the above into Eq. (\ref{eq:zt}) shows that $ZT$ of the
composite medium is always smaller than that of the high-$ZT$
material constituent. Fig. (\ref{fig:kappa}c) shows $ZT$ as a
function of concentration.  It is seen that under the favorable
parameter set considered here, the high $ZT$ value  is maintained
for a substantial amount of low $ZT$ material doping.  Generally
$ZT$ shows a decrease by percolation, and rises rapidly to the high
value just above this threshold.

I note that this model ignores interface scattering contributions to
$\sigma,\kappa$, and $L^{12}$.  If these contributions lower
$\kappa$ and $\sigma$ more drastically than $L^{12}$, it would imply
an overall enhancement in $ZT$ \cite{leonard}.  A more complete
theory requires a combination of bulk modeling as presented here,
combined with more microscopic modeling of materials interfaces.
Nevertheless the model demonstrates that disorder in the diffusive
regime can explain the enhancement in power factor, and can alter
the relationship between electrical and thermal conductivity of
electrons.  These are both important considerations in the
interpretation of experimental results, and in assessing the
viability of materials for improved thermoelectric efficiencies.

I gratefully acknowledge very helpful conversations with Fred
Sharifi, who introduced me to this problem.


\begin{thebibliography}{00}
\bibitem{footnote}
The equation of continuity for heat current is generally $\nabla
\cdot j_Q = j \cdot \nabla V + j_Q \cdot \nabla T/T$.  In linear
response the right-hand-side of this equation can generally be
ignored. The full nonlinear version with heating was solved in the
numerical calculation, and was found to agree with the results when
heating is neglected.
\end{thebibliography}
\end{document}